\documentclass[12pt,a4paper]{article}
\pdfoutput=1   
\usepackage[utf8]{inputenc}
\usepackage{amsmath,amsfonts,amssymb,amsbsy}
\usepackage{mathrsfs,latexsym}
\usepackage{graphicx}
\usepackage{color}
\usepackage{booktabs}
\usepackage{multirow}
\usepackage{multicol}
\usepackage{slashed}
\usepackage[table,xcdraw]{xcolor}
\usepackage{subfig}
\usepackage{alpha}
\hypersetup{
  pdftitle = {Charm loop effects on heavy meson masses and charmonium decay constants},
  pdfauthor = {S.Cali, K.Eckert, F.Knechtli, T.Korzec, J.Heitger},
  pdfkeywords = {Lattice QCD, Decoupling, Charm quark, Decay constants, $B_c$ mesons}
}%
\usepackage{newtxtext,newtxmath}

\newcommand{\bea}{\begin{eqnarray}}
\newcommand{\eea}{\end{eqnarray}}
\newcommand{\eq}[1]{Eq.~(\ref{#1})}

\newcommand{\msbar}{{\rm \overline{MS\kern-0.05em}\kern0.05em}}

\newcommand{\Nf}{{N_{\rm f}}}

\newcommand{\Ob}{\mathcal{O}}
\newcommand{\fP}{f_P}
\newcommand{\fV}{f_V}

\begin{document}

\preprintno{%
WUB/21-00 \\
MS-TP-21-04 
}

\title{%
Charm sea effects on charmonium decay constants and heavy meson masses}

\collaboration{\includegraphics[width=2.8cm]{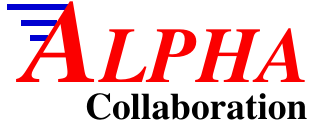}}

\author[mit,uj]{Salvatore~Cal{\`i}}
\author[mue]{Kevin~Eckert}
\author[mue]{Jochen~Heitger}
\author[wup]{Francesco~Knechtli}
\author[wup]{Tomasz~Korzec}
\address[mit]{Center for Theoretical Physics, Massachusetts Institute of Technology, Cambridge, MA 02139, USA}
\address[uj]{Institute of Theoretical Physics, Jagiellonian University, ul. \L ojasiewicza 11, 30-348 Krak\'ow, Poland}
\address[mue]{Institut für Theoretische Physik, Westfälische Wilhelms-Universität Münster, Wilhelm-Klemm-Straße~9,~48149 Münster, Germany}
\address[wup]{Department of Physics, Bergische Universit{\"a}t Wuppertal, Gau\ss str.~20, 42119~Wuppertal, Germany}

\begin{abstract}
\rule{\textwidth}{0.4pt}\\[\baselineskip] 
We estimate the effects on the decay constants of charmonium and on heavy meson masses due to the charm quark in the sea. 
Our goal is to understand whether for these quantities $\Nf=2+1$ lattice QCD simulations provide results that can 
be compared with experiments or whether $\Nf=2+1+1$ QCD including the charm quark in the sea needs to be simulated. 
We consider two theories, $\Nf=0$ QCD and QCD with $\Nf=2$ charm quarks in the sea. 
The charm sea effects (due to two charm quarks) are estimated comparing the results obtained in these two 
theories, after matching them and taking the continuum limit. The absence of light quarks allows us to simulate 
the $\Nf=2$ theory at lattice spacings down to $0.023$ fm that are crucial for reliable continuum extrapolations. 
We find that sea charm quark effects are below $1$\% for the decay constants of charmonium. Our results show that 
decoupling of charm works well up to energies of about $500\,$MeV. We also compute the derivatives of the decay 
constants and meson masses with respect to the charm mass. For these quantities we again do not see a significant dynamical charm quark effect, albeit with a lower precision. For mesons made of a charm quark and a heavy antiquark, whose mass is twice that of the charm quark, sea effects are only about $1${\textperthousand} in the ratio of vector to pseudoscalar masses.
\\  
\rule{\textwidth}{0.4pt}\\[\baselineskip] 
\end{abstract}
\vspace*{-0.5cm}

\maketitle 

\tableofcontents
\newpage
\section{Introduction}
In recent years there has been a renewed interest in spectral calculations of charmonium states, which are bound systems made of a charm quark $c$ and a charm antiquark $\bar{c}$. The motivation is due to the experimental discovery of many unexpected states, whose properties are still poorly understood \cite{Godfrey:2008nc}. Different hypotheses have been put forward, that see them identified as hybrid mesons, tetra-quarks or some other hitherto unknown form of matter. Other systems of particular interest are the $B_c$ mesons, which are heavy mesons composed of a bottom quark (antiquark) and a charm antiquark (quark).
Their peculiarities rely on the fact that they are the only heavy mesons consisting of heavy quarks with different flavors. Therefore, they cannot annihilate into gluons and they are more stable, having decay widths less than a hundred of keV. Only the pseudoscalar $B_c$ has so far been seen at the Tevatron collider at Fermilab \cite{Aaltonen:2007gv,Abazov:2008kv}. The high luminosity of the LHC collider at CERN now allows to measure the spectroscopy and the decay of $B_c^{\star}$ mesons with much better precision and therefore, comparisons of theoretical predictions with experiments will become more and more important. 

Lattice QCD may represent a powerful tool to understand the properties of charmonium and $B_c$ mesons directly starting from the QCD Lagrangian, for recent calculations see \cite{Liu:2012ze,Cheung:2016bym,DeTar:2018uko,Knechtli:2019bqx} (charmonium), \cite{Hatton:2020qhk} (charmonium decay constants) and \cite{McNeile:2012qf} ($B_c$ mesons). Depending on the problem at hand, numerical simulations usually include the lightest two, three or four flavors. Moreover, exact isospin symmetry is often assumed, implying that the lightest two quarks (up and down) are mass degenerate. The setups mentioned above are usually referred as  $\Nf=2$, $\Nf=2+1$ and $\Nf=2+1+1$ QCD. Even though lattice QCD simulations with $\Nf=2+1+1$ dynamical quarks would be highly desirable to have a better understanding of physical phenomena governed by the strong interaction, the addition of a dynamical charm quark to the sea typically complicates the generations of the gauge configurations on which observables are computed. Also, if lattice spacings are not fine enough, the results may be affected by large cutoff effects, owing to the heavy mass of the charm quark.

Given the present difficulties described above, in this work we want to quantify how much quenching the charm quark affects the results obtained through $\Nf=2+1$ QCD simulations (including mass degenerate up, down and a strange quarks). To this aim, we consider a model of QCD, namely QCD with two degenerate heavy quarks, whose mass is tuned to approximately reproduce the physical value of the charm quark mass. In order to quantify the impact of the dynamical heavy quarks, the results from this model are then compared to the ones found using $\Nf=0$ QCD (all quarks are quenched). The advantage of our strategy is that reliable continuum extrapolations can be performed, because light quarks are absent and we can use gauge ensembles with extremely fine lattice spacings ($a\lesssim 0.02$~fm) and relatively small volumes. In Refs.\cite{Korzec:2016eko,Cali:2018owe,Cali:2019enm} we followed this approach to evaluate the impact of a dynamical charm quark on the masses of charmonium states, e.g. on the pseudoscalar and vector meson masses $m_P$ and $m_V$.  From these studies it turned out that such effects are very small and for the hyperfine splitting $(m_V-m_P)/m_P$ they are of around $2\%$. 

Here, we want to extend our previous studies and investigate the charm quark sea effects on the hyperfine splitting of heavy mesons $H_c$ (bound states made of a heavy quark and a charm antiquark, or vice versa) and on the decay constants of charmonium. In particular, the focus is on the decay constants $\fP$ and $\fV$ of pseudoscalar and vector mesons, for which we follow the definitions of Ref.~\cite{Gasser:2010wz, Davies:2010ip}
\begin{eqnarray}
	\fP m_{P} &=& \langle 0\vert A_0^{q_1q_2}(0) \vert P(\vec{p}=0)\rangle,\label{eq:def_f_etac_cont}\\
	\fV m_{V} &=& \frac{1}{3}\sum_{i=1}^3\,\langle 0\vert V_i^{q_1q_2}(0) \vert V_i(\vec{p}=0)\rangle,
\label{eq:def_f_jpsi_cont}
\end{eqnarray}
where the bilinears containing quarks $q_1$ and $q_2$ are
\begin{equation}
   V^{q_1q_2}_\mu = \bar{q}_1 \gamma_\mu q_2, \qquad A^{q_1q_2}_\mu = \bar{q}_1 \gamma_\mu\gamma_5 q_2, \qquad P^{q_1q_2} = \bar{q}_1 \gamma_5 q_2,
\end{equation}
and $\vert P\rangle$ and $\vert V_i\rangle$ represent the ground state of a pseudoscalar and vector meson polarized in the direction $i$ of the vector current,\footnote{A definition of $\fV$ with general vector meson polarization can be found in Eq. (12) of \cite{Blossier:2018jol}.} respectively. The meson spatial momentum is $\vec{p}=0$ and $\vert 0 \rangle$ is the QCD vacuum state.\footnote{Note that we are using the convention for which the pion decay constant takes the value $f_{\pi}=130$ MeV.} 
In the charmonium system, the lowest-lying states lie below the $D\bar{D}$ threshold, resulting in relatively narrow widths due to the absence of OZI allowed \cite{Close:1979bt} strong decays. This means that radiative transitions, i.e. transitions from an initial state to a final state via the emission of a photon, can have significant experimentally accessible branching ratios. Therefore, the lattice calculation of the decay constants addressed here provides valuable theoretical insight for experiment on the fully non-perturbative level, such that we consider them as natural and representative observables to quantify charm sea quark effects in hadron physics beyond the mass spectrum.

The manuscript is organized as follows. In Section 2 we describe how decoupling works for observables which depend explicitely on the charm quark. In Section 3 we present our ensembles of gauge configurations with $\Nf=2$ charm quarks in the sea and $\Nf=0$ (pure gauge). The observables, from which we extract the decay constants and the heavy meson masses, are explained in Section 4, together with the tuning of the charm quark mass. The latter involves derivatives with respect to the heavy quark mass and we define suitable mass dependence functions in Section 4. Section 5 describes the distance preconditioning for the computation of the heavy quark propagator. The results of our calculations are presented in Section 6 and summarized in the conclusions (Section~7). Appendix~\ref{s:tabledecay} tabulates the values of the decay constants and their derivatives on our ensembles. Appendix~\ref{sec:etaX} shows a derivation of the formula for the mass derivative in the effective theory which accounts for the dependence of the scale of the effective theory on the mass of the heavy quarks.

\section{Decoupling in the presence of heavy valence quarks}

The decoupling of heavy quarks from low-energy physics is well 
understood~\cite{Weinberg:1980wa, Bernreuther:1981sg,Chetyrkin:2005ia,Schroder:2005hy} and 
is known to work very well already with heavy quarks as light as the charm quark~\cite{Athenodorou:2018wpk,Knechtli:2017xgy,Hollwieser:2020qri}.
QCD with light and a charm quark, described by the Euclidean action
\begin{equation}
   S_I = S^{\rm YM}[A_\mu] + S^{\rm light}[\bar \psi_{u,d,s}, \psi_{u,d,s},A_\mu] + (\bar \psi_c, [\slashed D[A_\mu] + m_c] \psi_c)\, ,
\end{equation}
where $(\ldots,\ldots)$ denotes the standard inner product for spinors, can be approximated by an effective theory, which to leading order is given by QCD without the heavy quarks
\begin{equation}
   S_I^{\rm dec} = S^{\rm YM}[A_\mu] + S^{\rm light}[\bar \psi_{u,d,s}, \psi_{u,d,s},A_\mu]\, .
\end{equation}
When the couplings of the effective theory are correctly ``matched'', they inherit a dependence on the heavy quark mass of the fundamental theory, and the effective theory reproduces
results of the fundamental one, as long as the observables do not contain heavy quarks and are 
low-energy quantities. The differences between the two theories are of order $O((\Lambda/\overline m_c)^2)$,
$O((\overline m_{\rm light}/\overline m_c)^2)$ and $O((E/\overline m_c)^2))$, where $E$ is the energy scale 
associated with the observable at hand, e.g. a momentum transfer in a process involving the light quarks, and $\overline m_{c,\rm light}$ are the running masses of light and charm quarks. 
These differences could in principle be suppressed further by adding additional terms to the 
effective action, e.g.
\begin{equation}
   \frac{c}{m_c^2} (\bar \psi_{\rm u,d,s}, \psi_{\rm u,d,s})^2\, .
\end{equation}

It is common practice in lattice QCD, to simulate one of the effective theories instead of 
full QCD. In particular, $b$ and $t$ quarks are never included in the simulations while the
$c$ quark sometimes is included and sometimes not. Even without $c$ quarks in the action, 
charm physics is studied in the ``partially quenched'' approximation. Before we study how well this approximation works numerically, we would like to argue that it is to some extent 
covered by the theory of decoupling, even though the observables now {\em do} depend
on the charm quarks. We can add two terms to $S_I$ that amount to a factor of $1$ in the
path integral
\begin{equation}
   S_{II} = S_{I} + (\bar \psi_{c'}, [\slashed D[A_\mu] + m_c] \psi_{c'}) + (\phi^\dagger_{c'}, [\slashed D[A_\mu] + m_c] \phi_{c'})\, .
\end{equation}
While $\psi_{c'}$ is a Grassmann-valued fermion field, $\phi$ is a complex-valued ``pseudo-fermion'' field with 
the same space-time, color and spin components.
After integration over the fermion and pseudo-fermion fields, the fermionic determinant is exactly cancelled by the 
determinant from the pseudo-fermion Gaussian 
integral.\footnote{This argument can be made more rigorous on the lattice, e.g. with Wilson fermions, and depends on 
$m_c$ being large enough for the Wilson operator to only have eigenvalues with positive real parts.} Therefore, expectation values in the two theories are equal
\begin{eqnarray}
	\langle \Ob[\bar\psi_{u,d,s}, \psi_{u,d,s}, \bar \psi_c, \psi_c, A_\mu]\rangle_I &=& 
	\langle \Ob[\bar\psi_{u,d,s}, \psi_{u,d,s}, \bar \psi_c, \psi_c, A_\mu]\rangle_{II}\nonumber \\
	&=& \langle \Ob[\bar\psi_{u,d,s}, \psi_{u,d,s}, \bar \psi_{c'}, \psi_{c'}, A_\mu]\rangle_{II}\, .
\end{eqnarray}
The decoupling mechanism can now be applied to the $\psi_c$ fields in $S_{II}$, leading to a leading-order effective theory
\begin{equation}
   S_{II}^{\rm dec} = S_{I}^{\rm dec} + (\bar \psi_{c'}, [\slashed D[A_\mu] + m_c] \psi_{c'}) + (\phi^\dagger_{c'}, [\slashed D[A_\mu] + m_c] \phi_{c'})\, .
\end{equation}
After matching of the parameters (gauge coupling, light quark masses and the mass of the ``partially quenched'' charm quark), one expects
\begin{equation}
   \langle \Ob[\bar\psi_{u,d,s}, \psi_{u,d,s}, \bar \psi_{c'}, \psi_{c'}, A_\mu]\rangle_{II}
   = \langle \Ob[\bar\psi_{u,d,s}, \psi_{u,d,s}, \bar \psi_{c'}, \psi_{c'}, A_\mu]\rangle_{II}^{\rm dec} + O(1/m_c^2)\, .
\end{equation}
With the mass of the $c'$ quark being the same as the one that was integrated out, one can of course expect large $O(1)$ corrections,
but in some cases the energy scale of the observable may still be low, even if it contains heavy quarks.
For example, the binding energies in charmonia are not very high, compared to the charm quark mass. 
In the same respect, another case are their decay constants which turn out to be not very large.

\section{Numerical setup}\label{sec:numerical-setup}
Our calculations are performed on the gauge ensembles with parameters summarized in Table~\ref{tab:ensembles}.
\begin{table}[h!]
\begin{center}
{\footnotesize
\renewcommand{\arraystretch}{1.4}
\renewcommand{\tabcolsep}{5.5pt}
\begin{tabular}{c | c c c c c c c c c}
\toprule
$\Nf$ & ID & $\frac{T}{a}\times\left(\frac{L}{a}\right)^3$ &  $\beta$  & $\kappa$    & $a \mu$            & $\sqrt{t_0}m_{P}$ & $t_0/a^2$ & $a$ [fm] & MDUs\\
\midrule
2 & E &$95\times 24^3$  &  5.300  & 0.135943    & 0.36151  & 1.79303(55)      & 1.23907(82) & 0.104 & 8000\\
 & N & $119\times 32^3$                              &  5.500  & 0.136638    &  0.165997          & 1.8048(15)      & 4.4730(93) & 0.054 & 8000\\
 & P & $119\times 32^3$                              &  5.700  & 0.136698    & 0.113200           & 1.7931(28)     & 9.105(35) & 0.038 & 17184\\
 & S & $191\times48^3$                               &  5.880  & 0.136509    & 0.087626           & 1.8130(29)     & 15.621(60) & 0.029 & 23088\\
 & W & $191\times 48^3$                              &  6.000  & 0.136335    & 0.072557           & 1.8075(43)     &22.39(12) & 0.024 &  22400\\
\midrule
0 & qN & $119\times 32^3$                               &  6.100  &    --       &    0.16              & 1.69807(67)  & 4.4329(38) & 0.054 & 64000 \\
 &  &  &  &  &  0.17  & 1.76547(70)  &  &  &  \\
 &  &  &  &  &  0.18  & 1.83195(72)  &  &  &  \\
 & qP & $119\times 32^3$                               & 6.340  &    --       &    0.11              & 1.6856(22)  & 9.037(30) & 0.038 & 20080\\
 &  &  &  &  &  0.12  & 1.7848(24)  &  &  &  \\
 &  &  &  &  &  0.13  & 1.8824(25)  &  &  &  \\

 & qW & $191\times 48^3$                               &  6.672  &    --       &    0.07              & 1.6830(28)  & 21.925(83) & 0.024 & 73920\\
  &  &  &  &  &  0.08  & 1.8399(30)  &  &  &  \\
 &  &  &  &  &   0.09  & 1.9934(33)  &  &  &  \\

 & qX & $191\times 64^3$                               & 6.900  &    --       &    0.056              & 1.7714(28)  & 39.41(14) & 0.018 & 160200\\
 &  &  &  &  &  0.058  & 1.8137(29)  &  &  &  \\
 &  &  &  &  &  0.060  & 1.8558(30)  &  &  &  \\

\bottomrule
\end{tabular}
}
\end{center}
\caption[Simulation parameters of our ensembles.]{Simulation parameters of our ensembles. The columns show the lattice sizes, 
the gauge coupling $\beta=6/{g_0^2}$, the critical hopping parameter, the twisted mass parameter $\mu$, the pseudoscalar mass in $t_0$ units, 
the hadronic scale $t_0/a^2$ defined in \cite{Luscher:2010iy}, the lattice spacing in fm from the scale $t_0$ \cite{Cali:2019enm} and the total statistics in molecular dynamics units. For the $\Nf=0$ ensembles, we report the values of $\sqrt{t_0}m_{P}$ measured at three different valence quark masses $a\mu$.
The temporal extent $T$ is an odd multiple of $a$ because open boundary conditions are applied in the temporal directions and the links pointing out of the lattice volume are absent.
}\label{tab:ensembles}
\end{table}
The $\Nf=0$ ensembles are generated using the standard Wilson plaquette gauge action~\cite{Wilson:1974sk}, whilst for the $\Nf=2$ case a clover-improved~\cite{Sheikholeslami:1985ij} doublet of twisted mass Wilson fermions~\cite{Frezzotti:2000nk,Frezzotti:2003ni} is added. 

Since we aim at performing continuum extrapolations using very fine lattice spacings, open boundary conditions in the time direction are applied~\cite{Luscher:2011kk} to avoid the well known problems related to the deficient sampling of topological sectors. The spatial dimensions are kept periodic. Moreover, for the production of our ensembles we benefit from the knowledge of the critical mass $m_{cr}$~\cite{Fritzsch:2012wq,Fritzsch:2015eka} and the axial current and pseudoscalar density renormalization factors $Z_A$~\cite{Luscher:1996jn,DellaMorte:2005xgj,DallaBrida:2018tpn} and $Z_P$~\cite{Fritzsch:2012wq,Juttner:2004tb}. For further details about the simulations we refer to our previous works~\cite{Knechtli:2017xgy,Cali:2019enm} and references therein.

\section{Observables}
\subsection{Computation of meson decay constants with open boundary conditions}\label{subsec:decay_constants}
Meson decay constants are related to matrix elements between the vacuum and the meson state at rest, as anticipated in Eqs.~\eqref{eq:def_f_etac_cont} and~\eqref{eq:def_f_jpsi_cont}. In the framework of the model considered here, we denote the doublet of the heavy degenerate charm quarks in the physical basis as $\psi=(\tilde{c}_1,\tilde{c}_2)^{\intercal}$ and its counterpart in the twisted basis as $\chi=(c_1,c_2)^{\intercal}$. At maximal twist, the relation between physical and twisted basis is
\begin{equation}
\psi = \frac{1+i\gamma_5\tau^3}{\sqrt{2}}\chi,\qquad \bar{\psi}=\bar{\chi}\frac{1+i\gamma_5\tau^3}{\sqrt{2}}.
\end{equation}
Therefore, the decay constants of pseudoscalar ($P$) and vector ($V$) mesons in twisted mass QCD are given by
\begin{eqnarray}
\fP m^2_{P} &=& 2\mu\,\langle 0\vert \bar{c}_1\gamma_5 c_2\vert P\rangle,\label{eq:def_f_etac}\\
\fV m_{V} &=& \frac{1}{3}\sum_{i=1}^3\,\langle 0\vert\bar{c}_1\gamma_i\gamma_5c_2\vert V_i\rangle,
\label{eq:def_f_jpsi}
\end{eqnarray}
where $\mu$ is the twisted mass parameter. To derive Eq.~\eqref{eq:def_f_etac}, PCVC relations in the twisted basis have been used, as described in Refs.~\cite{Jansen:2003ir,Cali:2019gqf}. Since both our quarks, $c_1$ and $c_2$, have the mass of a charm quark, $f_{P,V}$ and $m_{P,V}$ are comparable to the decay constants and masses of the charmonia states $\eta_c$ and $J/\psi$. The main differences are related to the fact that in our calculations we neglect the contribution of disconnected diagrams and we are assuming a model of QCD without light quarks and with two charm quarks instead of one.

The twisted mass formulation of QCD provides a particularly convenient setup for the calculation of the pseudoscalar decay constant $\fP$, because it simplifies the renormalization properties of our operators. In fact, the renormalization factors $Z_P$ and $Z_{\mu}$ obey the relation $Z_PZ_{\mu} = 1$. Thus, the lattice calculation of $\fP$ does not need any renormalization factors, as already discussed in Refs.~\cite{Frezzotti:2000nk,Jansen:2003ir}. As concerns the twisted mass expression for $\fV$, the matrix element in Eq.~\eqref{eq:def_f_jpsi} must be multiplied by the renormalization factor $Z_A$ of the axial current, which is known from Refs.~\cite{Luscher:1996jn,DellaMorte:2005xgj,DallaBrida:2018tpn} for the ensembles considered here.

To extract the decay constants $\fP$ and $\fV$ given in Eqs.~\eqref{eq:def_f_etac} and \eqref{eq:def_f_jpsi}, we first need to know the values of the masses $m_P$ and $m_V$ (already computed in Ref.~\cite{Cali:2019enm}) and of the matrix elements $\langle 0\vert \bar{c}_1\gamma_5 c_2\vert P\rangle$ and $\langle 0\vert\bar{c}_1\gamma_i\gamma_5c_2\vert V_i\rangle$. All these quantities can be determined from the lattice calculation of zero-momentum correlation functions of the form
\begin{eqnarray}
a^3 f_{PP}(x_0,y_0) &=& -\frac{a^3}{L^3}\sum_{\vec{x},\vec{y}} a^6
\langle P^{c_1c_2}(x_0,\mathbf{x})P^{c_2c_1}(y_0,\mathbf{y})\rangle,  \label{eq:def_fpp}\\
a^3 f_{A_iA_i}(x_0,y_0) &=& -\frac{a^3}{L^3}\sum_{\vec{x},\vec{y}} a^6
\langle A_i^{c_1c_2}(x_0,\mathbf{x})A_i^{c_2c_1}(y_0,\mathbf{y})\rangle, \label{eq:def_faiai}
\end{eqnarray}
where $(y_0,\mathbf{y})$ and $(x_0,\mathbf{x})$ denote the coordinates at which a particle state is created (source) and annihilated (sink), respectively. Integrating over the fermion fields, one obtains
\begin{equation}
\begin{split}
&a^8 \sum_{\vec{x},\vec{y}} \langle \bar{c}_1(x)\Gamma c_2(x)\bar{c}_2(y)\Gamma c_1(y)\rangle_{F}=-\sum_{\vec{x},\vec{y}}\text{Tr}\left[\Gamma D^{-1}_{c_2}(x, y)\Gamma D^{-1}_{c_1}(y, x)\right],\\
&\Gamma=\lbrace{\gamma_5,\gamma_i\gamma_5}\rbrace,
\label{eq:connected_piece}
\end{split}
\end{equation}
where $\langle \ldots\rangle_F$ is the expectation value over the fermion fields and $D^{-1}_{c1}$, $D^{-1}_{c_2}$ are the fermion propagators of the quark $c_{1}$ and $c_2$, respectively.
The trace in Eq.~\eqref{eq:connected_piece} can be efficiently estimated stochastically. In particular, we use time-dilution with 16 $U(1)$ noise sources per time-slice, which amounts to 16 inversions
per $y_0$ value and Dirac structure $\Gamma$. In principle, the projection to zero-momentum can be realized performing a sum either over $\vec{x}$ or $\vec{y}$. However, here we prefer to keep the two sums, as indicated in Eqs.~\eqref{eq:def_fpp} and ~\eqref{eq:def_faiai}, because that typically leads to highly improved signals for the meson correlators.

The pseudoscalar and vector meson masses are extracted from the exponential decay of the correlators \eqref{eq:def_fpp} and \eqref{eq:def_faiai}, respectively. For this purpose, we first compute the effective mass
\begin{equation}
am^{eff}(x_0+a/2,y_0) \equiv \ln\left(\frac{f(x_0,y_0)}{f(x_0+a,y_0)}\right)
\label{eq:effective_mass}
\end{equation}
and then perform a weighted average in the plateau region as
\begin{equation}
am = \frac{\sum\limits_{x_0/a=t_i}^{t_f}w(x_0+a/2,y_0)am_{eff}(x_0+a/2,y_0)}{\sum\limits_{x_0/a=t_i}^{t_f}w(x_0+a/2,y_0)},
\label{eq:eff_mass_plateau_average}
\end{equation}
where $x_0/a=t_{i,f}$ are the initial and final time-slices of the plateau, chosen such that the results do not depend on source position $y_0$ and excited state contributions are sufficiently suppressed. The weights $w$ are given by the inverse squared errors of the corresponding effective masses.

In principle, even the matrix elements $\langle 0\vert \bar{c}_1\gamma_5 c_2\vert P\rangle$ and $\langle 0\vert\bar{c}_1\gamma_i\gamma_5c_2\vert V\rangle$ can be extracted through an exponential fit to the meson correlators \eqref{eq:def_fpp} and \eqref{eq:def_faiai}. However, since we use open boundary conditions in the time direction, such an approach may lead to unreliable results, because of boundary effects. For this reason, we proceed along the method described in Refs.~\cite{Bruno:2016plf,Bruno:2016avt}, whose advantage is to remove the unwanted boundary effects  by forming a suitable ratio of two-point correlation functions. 

\begin{figure}[t]
\centering
\includegraphics[width=0.8\linewidth]{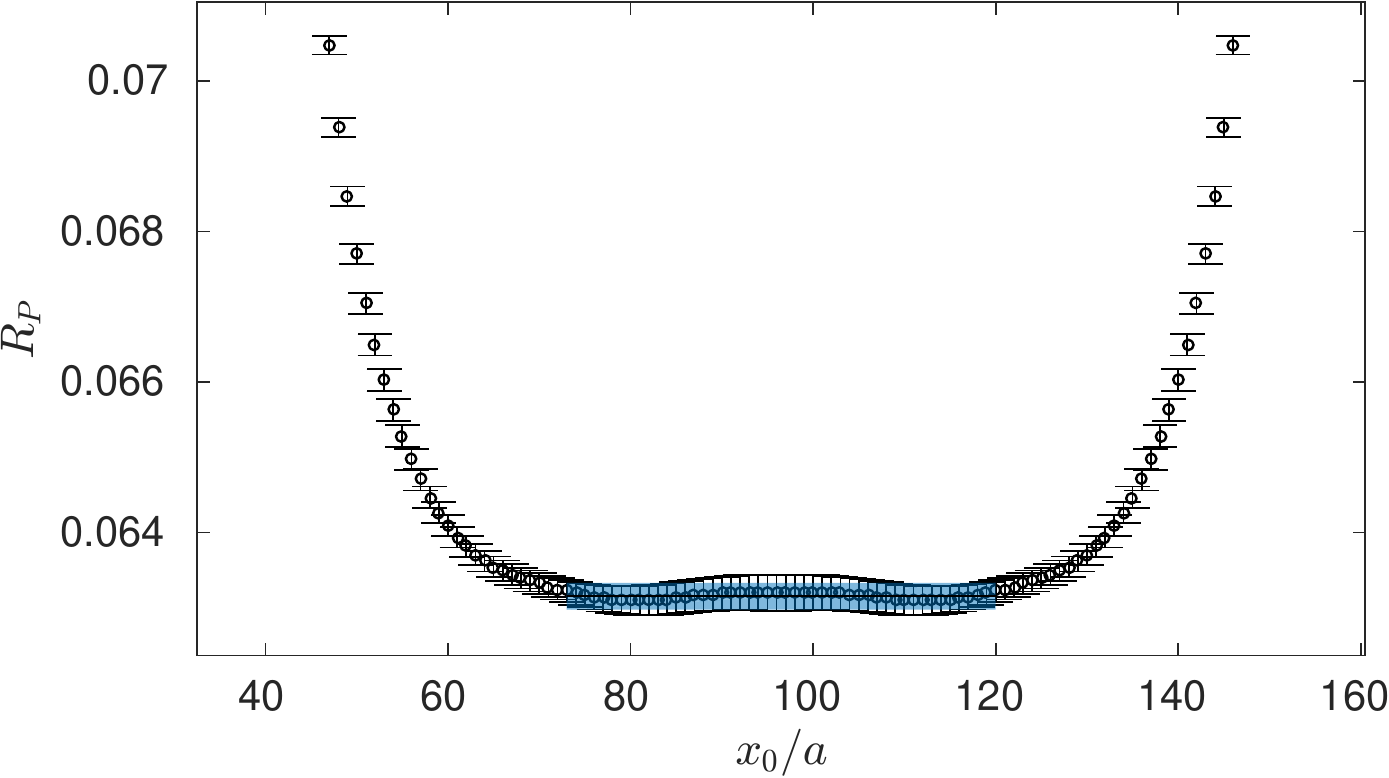}
 \caption{Effective quantity $R_P$, defined in \eqref{eq:rp}, from which the decay constants are extracted. The results shown in the figure are obtained on the $\Nf=2$ ensemble W at $\beta=6.0$. The horizontal error band indicates the plateau average $\bar{R}_{P}$ with its statistical uncertainty.} \label{fig:tmWm2_rp}
\end{figure}

Following the procedure of Ref.~\cite{Bruno:2016avt}, when $x_0$ is far enough from the boundaries $(0\ll x_0 \ll T )$ and $x_0-y_0$ is sufficiently large, it is possible to show that the correlators $f_{PP}$ and $f_{A_i A_i}$ have a leading asymptotic behavior according to\footnote{The usual relativistic normalization of states in a finite volume is used, i.e. $\langle 0|0\rangle=1$, $\langle P|P\rangle = 2m_P L^3$ and $\langle V_i | V_i\rangle = 2m_V L^3$.}
\begin{eqnarray}
\sqrt{2m_PL^3}f_{PP}(x_0,y_0) & \simeq &\langle 0\vert \bar{c_1}\gamma_5 c_2\vert P\rangle e^{-m_P (x_0-y_0)}A(y_0),\label{eq:asympt_fpp1}\\
\sqrt{2m_PL^3}f_{PP}(T-x_0,y_0) & \simeq &\langle 0\vert \bar{c_1}\gamma_5 c_2\vert P\rangle e^{-m_P (T-x_0-y_0)}A(y_0),\label{eq:asympt_fpp2}\\
L^3f_{PP}(T-y_0,y_0)& \simeq & e^{-m_P (T-2y_0)}A^2(y_0),\label{eq:asympt_fpp3}\\
\sqrt{2m_VL^3}f_{A_iA_i}(x_0,y_0)& \simeq & \langle 0\vert \bar{c_1}\gamma_i\gamma_5 c_2\vert V_i\rangle e^{-m_V (x_0-y_0)}B(y_0),\label{eq:asympt_faiai1}\\
\sqrt{2m_VL^3}f_{A_iA_i}(T-x_0,y_0)& \simeq & \langle 0\vert \bar{c_1}\gamma_i\gamma_5 c_2\vert V_i\rangle e^{-m_V (T-x_0-y_0)}B(y_0),\label{eq:asympt_faiai2}\\
L^3f_{A_i A_i}(T-y_0,y_0)& \simeq & e^{-m_V (T-2y_0)}B^2(y_0),\label{eq:asympt_faiai3}
\end{eqnarray}
where $A$, $B$ are dimensionless factors depending on the source position $y_0$ and related to the matrix elements between the vacuum and the boundary state. From the relations~\eqref{eq:asympt_fpp1}-\eqref{eq:asympt_faiai3} one reads off that the relevant amplitudes can be determined through the following ratios of correlators
\begin{eqnarray}
R_{P}(x_0,y_0)&\equiv&\sqrt{\frac{a^3\vert f_{PP}(x_0,y_0)f_{PP}(T-x_0,y_0)\vert}{f_{PP}(T-y_0,y_0)}} = \frac{a^2\langle 0\vert \bar{c}_1\gamma_5 c_2\vert P\rangle}{\sqrt{2am_{P}}}\label{eq:rp},\\
R_{A_i}(x_0,y_0)&\equiv&\sqrt{\frac{a^3\vert f_{A_iA_i}(x_0,y_0)f_{A_iA_i}(T-x_0,y_0)\vert}{f_{A_iA_i}(T-y_0,y_0)}} = \frac{a^2\langle 0\vert \bar{c}_1\gamma_i\gamma_5 c_2\vert V_i\rangle}{\sqrt{2a m_{V}}}\label{eq:ra}.
\end{eqnarray}
Employing such ratios, the matrix elements of operators close to the boundary drop out, so there are no restrictions in the choice of the source position $y_0$.  Eqs.~\eqref{eq:rp} and \eqref{eq:ra} are satisfied for a large range of sink positions $0\ll x_0\ll T$, where boundary effects and excited state contributions can be neglected. Thus, to improve the estimate of the matrix elements, we take the plateau averages
\begin{equation}
\bar{R}_{P}\equiv\frac{1}{t_f-t_i+1}\sum_{x_0/a=t_i}^{t_f}R_P(x_0,y_0),\qquad \bar{R}_{A_i}\equiv\frac{1}{t_f-t_i+1}\sum_{x_0/a=t_i}^{t_f}R_{A_i}(x_0,y_0),
\label{eq:Rbar_definition}
\end{equation}
where $t_i$ and $t_f$ are the start and the end of the plateau.
In Figure~\ref{fig:tmWm2_rp} we report the measurement of the effective quantity $R_P$ in Eq.~\eqref{eq:rp} obtained on the $\Nf=2$ ensemble~$W$, with parameters reported in Table~\ref{tab:ensembles}. As can be seen, the use of lattices with large temporal extent and very fine lattice spacings allows us to take the plateau average $\bar{R}_{P}$ for a large range of temporal slices. Similar conclusions also hold for the other ensembles and the ratios $R_{A_i}$.

Finally, combining the formulae above, the pseudoscalar and vector decay constants $\fP$ and $\fV$ in lattice units can be determined as
\begin{eqnarray}
a\fP &=& 2a\mu \bar{R}_{P}\sqrt{\frac{2}{(am_{P})^3}},\label{eq:fp_formula}\\
a\fV &=& Z_A\left(\frac{1}{3}\sum_{i=1}^3\bar{R}_{A_i}\right)\sqrt{\frac{2}{a m_{V}}}\label{eq:fv_formula}.
\end{eqnarray}
The crucial point is to obtain a reliable estimate of the correlators $f_{PP}(T - y_0 , y_0 )$ and
$f_{A_iA_i} (T - y_0 , y_0 )$ in Eqs.~\eqref{eq:rp} and \eqref{eq:ra}, as when dealing with heavy quarks the relative precision of the solution of the Dirac equation deteriorates at large distances. A detailed discussion of this issue will be given in Section~\ref{sec:distance_preconditioning}. To improve the quality of the signal, we average over forward and backward correlator. Therefore, we consider the symmetrized correlators
\begin{eqnarray}
\bar{f}_{PP}(x_0,y_0) &=& \frac{1}{2}\left[ f_{PP}(x_0,y_0) + f_{PP}(T-x_0,T-y_0)\right],\\
\bar{f}_{A_iA_i}(x_0,y_0) &=& \frac{1}{2}\left[ f_{A_iA_i}(x_0,y_0) + f_{A_iA_i}(T-x_0,T-y_0)\right].
\end{eqnarray}
As it was demonstrated in Appendix~A of Ref.~\cite{Cali:2019enm}, taking the average with the time-reflected correlators prohibites the mixing with states of opposite parity.\footnote{Parity is not a symmetry in the twisted mass formulation of QCD for twisted mass parameter $\mu\neq0$.}

\subsection{Heavy mesons}\label{sec:obs_heavy_mesons}
To study the charm loop effects at even higher energies, we measure the masses of heavy mesons $H_c$ made of a heavy quark, $h$, with mass $m_h= \lbrace m_c, 2m_c\rbrace$. In particular, we focus on the ground state of the pseudoscalar and vector channels. The masses are extracted according to Eqs.~\eqref{eq:effective_mass},~\eqref{eq:eff_mass_plateau_average}. The main  ingredient of the calculation is the propagator $D_{h}^{-1}$, which is evaluated using the twisted mass parameter $\mu_h= \mu$ or $\mu_h= 2\mu$, depending on the value of the mass $m_h$ we want to impose. Let us recall that $\mu$ is the twisted mass parameter of the simulations tuned to reproduce the charm quark mass.

\begin{figure}[t]
\centering
\includegraphics[width=0.8\linewidth]{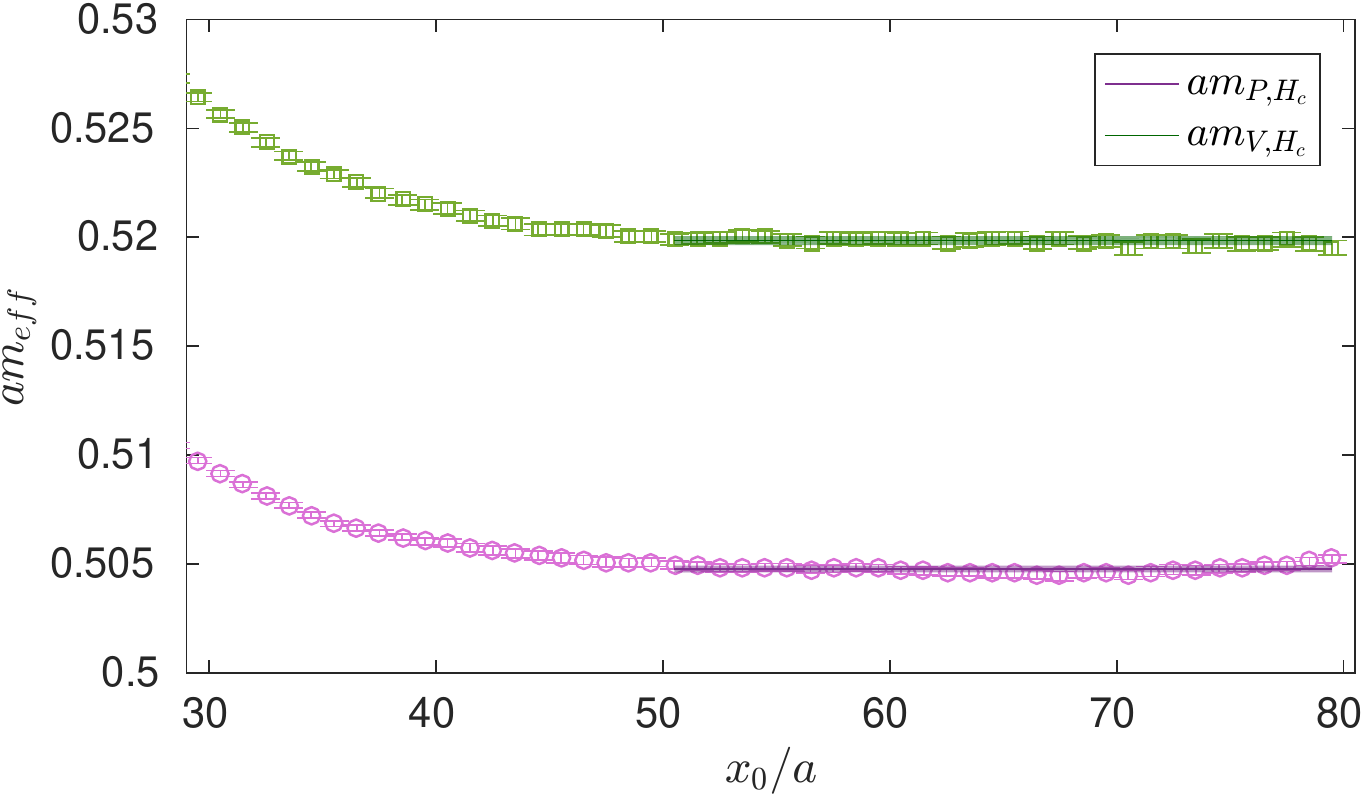}
 \caption{The effective masses for heavy mesons made of a heavy quark $h$ ($m_h=2m_c$) and a charm quark $c$ on the W ensemble are displayed, together with the plateau average and its error band. We show the pseudoscalar (violet circles) and vector (green squares) channels.}\label{fig:eff_mass}
\end{figure}

From now on, we refer to the masses of these heavy mesons using the notation
\begin{equation}
m_{X,H_c},\quad X=\lbrace P,V\rbrace ,
\end{equation}
where $X$ denotes the vector or pseudoscalar state and the heavy quark, $h$, can have a mass equal to $m_c$ or $2m_c$.
The masses are extracted through the plateau average \eqref{eq:eff_mass_plateau_average} of the effective mass \eqref{eq:effective_mass}.
For the $m_h=m_c$ case, we refer to our previous work~\cite{Cali:2019enm}.
An example of this procedure for $m_h=2m_c$ is shown in Figure \ref{fig:eff_mass} for the $\Nf=2$ ensemble $W$.

\subsection{Tuning of the twisted mass parameter}
To match $\Nf=0$ and $\Nf=2$ QCD, we use the low-energy observable $m^{had}=1/\sqrt{t_0}$. For such observables decoupling applies \cite{Appelquist:1974tg,Weinberg:1980wa} and we can assume $\sqrt{t_0}\vert_{\Nf=0} = \sqrt{t_0(M_c)}\vert_{\Nf=2}$. In order to compare the two theories, the value of the renormalization group invariant (RGI) mass $M_c$ of the charm quark needs to be fixed and, following the procedure of our previous work \cite{Cali:2019enm}, we choose $M_c$ such that $\sqrt{t_0}m_{P}$ is approximately equal to the physical value for the $\eta_c$ meson and is the same in $\Nf=0$ and $\Nf=2$ QCD, for each lattice spacing employed (see Table \ref{tab:ensembles}).

To set $M_c$ we proceed as described below. On our finest $\Nf=2$ lattice $(\beta=6.0)$, the RGI mass $M_c$ of the charm quark is fixed by the relation 
\begin{equation}
M_c/\Lambda_{\msbar}=4.87,
\label{eq:rgi_mass_mc}
\end{equation}
where we use the value $M_c = 1510$ MeV of Ref.~\cite{Heitger:2013oaa} (which agrees with Ref.~\cite{Tanabashi:2018oca}, see also Ref.~\cite{Heitger:2021apz}) and the two-flavor Lambda parameter $\Lambda_{\msbar}=310(20)$ MeV known from \cite{Fritzsch:2012wq}. Since on our $\Nf=2$ ensembles the hopping parameter $\kappa$ is set to its critical value, the renormalized physical quark mass is simply given by $m_R=Z_P^{-1}\mu$ and the condition \eqref{eq:rgi_mass_mc} is equivalent to fix the twisted mass parameter through
\begin{equation}
a\mu =\frac{M_c}{\Lambda_{\msbar}}\times Z_P(L_1^{-1}) \times \frac{m_R(L_1)}{M_c} \times \Lambda_{\msbar}L_1 \times \frac{a}{L_1}.
\label{eq:set_mu}
\end{equation}
In the equation above, the value of the pseudoscalar renormalization factor $Z_P$ at the renormalization scale $L_1^{-1}$ in the Schr\"odinger Functional scheme
\begin{equation}
Z_P(L_1)=0.5184(33), \quad\mbox{valid for } 5.2\leq\beta\leq 6,
\label{eq:ZP_L1}
\end{equation}
and the relation between the running mass $m_R(L_1)$ and the RGI mass\
\begin{equation}
\frac{M}{m_R(L_1)}=1.308(16)
\label{eq:M_over_mR}
\end{equation}
are available from Refs.~\cite{Fritzsch:2012wq,DellaMorte:2005kg}. As a value of the $\Lambda$ parameter of the two-flavor theory in $L_1$ units we take~\cite{DellaMorte:2004bc}
\begin{equation}
\Lambda_{\msbar} L_1=0.649(45),
\label{eq:Lamda_L1}
\end{equation}
whilst the ratio $L_1/a$ is known from~\cite{Athenodorou:2018wpk}. Plugging the values of Eqs.~\eqref{eq:ZP_L1}-\eqref{eq:Lamda_L1} into Eq.~\eqref{eq:set_mu}, the charm quark mass can be fixed at a precision of around $10\%$. Such accuracy is acceptable for us if we devise a method to keep the relative mass differences between the different ensembles under better control. In practice, instead of using  Eq.~\eqref{eq:set_mu}, we tune the twisted mass parameter to a point $\mu^{\star}$ such that the renormalized quantity $\sqrt{t_0}m_{\eta_c}$ is equal to its value on our finest $\Nf=2$ ensemble W at $\beta=6.0$ (see Table \ref{tab:ensembles}) and satisfies
\begin{equation}
\sqrt{t_0}m_{P}(\mu^{\star}) \equiv 1.807463 \,.
\label{eq:tuning_point}
\end{equation}
This value is obtained on the ensemble W by setting the twisted mass parameter through Eq.~\eqref{eq:set_mu}.
The tuning point Eq.~\eqref{eq:tuning_point} is independent of the overall scale $\Lambda_{\msbar}$ (known with a $7\%$ accuracy and which limits the overall accuracy).

In the $\Nf=2$ theory, we apply a correction to all observables, which is based on the computation of twisted mass derivatives \cite{Cali:2019enm}. First, we determine the target tuning point $\mu^{\star}$ through the Taylor expansion
\begin{equation}
\mu^{\star} = \mu + (\sqrt{t_0}m_{P} - 1.807463)\left(\frac{d\sqrt{t_0}m_{\eta_c}}{d\mu}\right)^{-1} \,,
\label{eq:mustar}
\end{equation}
where $\mu$ and $\sqrt{t_0}m_{P}$ are the values from the simulations
obtained using Eq.~\eqref{eq:set_mu}, see Table \ref{tab:ensembles}.
Subsequently all quantities, denoted by $A$ below, are corrected by
\begin{equation}
A(\mu^{\star}) = A(\mu) + (\mu^{\star} - \mu)\frac{dA}{d\mu}.
\label{eq:phistar}
\end{equation}
For a generic primary observable $A$, its twisted mass derivative is simply given by
\begin{equation}
\frac{d\langle A\rangle}{d\mu} = -\left\langle\frac{dS}{d\mu}A\right\rangle + \left\langle\frac{dS}{d\mu}\right\rangle\langle A\rangle + \left\langle\frac{dA}{d\mu}\right\rangle,
\label{eq:dAdmu}
\end{equation}
where $S$ is the lattice QCD action. However, most quantities we are interested in are non-linear functions $f$ of various primary observables, for which we apply the chain rule 
\begin{equation}
\frac{df(\langle A_1\rangle,\dots,\langle A_N\rangle,\mu)}{d\mu} = \frac{\partial f}{\partial \mu} + \sum_{i=1}^N\frac{\partial f}{\partial\langle A_i\rangle}\frac{d\langle A_i\rangle}{d\mu}.
\label{eq:dfdmu}
\end{equation}
After integrating over the fermion fields, Eq.~\eqref{eq:dAdmu} becomes
\begin{equation}
  \frac{d\langle \tilde{A}\rangle^{\rm gauge}}{d\mu} =
-\left\langle\frac{dS_{\rm eff}}{d\mu}\tilde{A}\right\rangle^{\rm gauge} +
\left\langle\frac{dS_{\rm eff}}{d\mu}\right\rangle^{\rm gauge}\langle \tilde{A}\rangle^{\rm gauge}
+ \left\langle\frac{d\tilde{A}}{d\mu}\right\rangle^{\rm gauge} \,,
\label{eq:dAdmu_gauge}
\end{equation}
where $\langle \cdot \rangle^{\rm gauge}$ are expectation values taken on the
ensemble of gauge fields and $S_{\rm eff}$ is the effective lattice QCD gauge
action which includes the effect of the fermion determinants. The observable
$\tilde{A}$ contains in general the fermion propagator $D^{-1}$. The sea quark
effects originate from $\frac{dS_{\rm eff}}{d\mu}$. They affect the mass
derivative of an observable through the covariance of $\frac{dS_{\rm eff}}{d\mu}$ and $\tilde{A}$.

In the simplest case of $\Nf=0$ QCD, the action does not depend on the quark masses and the twisted mass parameter $\mu$ only enters the inversions of the Dirac operator. The mass derivative of an observable in the $\Nf=0$ theory therefore only receives a contribution from the last term on the right hand side of Eq.~\eqref{eq:dAdmu_gauge}. Thus, to reproduce the tuning point $\mu^{\star}$, Eq.~\eqref{eq:tuning_point}, for our $\Nf = 0$ ensembles, we measure the quantities of interest at three different values of the twisted mass parameter $\mu$ such that $\mu^{\star}$ is eventually obtained through a linear interpolation of the measurements. 

\subsection{Mass dependence functions}
The mass dependence of a  multiplicatively renormalizable quantity $X$ is captured by the renormalized combination
\begin{equation}
   \eta_X \equiv \frac{M}{X} \frac{dX}{dM}\, .\label{eq:etaX}
\end{equation}
Interesting quantities $X$ are for instance the vector or pseudoscalar meson masses as studied in~\cite{Cali:2019enm}, or their decay constants, as
determined in this work. In the twisted mass formulation at maximal twist, the RGI quark mass is connected multiplicatively to the twisted mass
parameter $\mu$, which implies that $M$ can be replaced by $\mu$ in Eq.~\eqref{eq:etaX}. When computed in the effective theory, the mass dependence
stems from an explicit mass dependence of the quantity in question, but also from the mass dependence that the scale inherits through the matching
condition between the effective and full theory. In this context it is useful to also define
\begin{equation}
   \tilde \eta_X \equiv \frac{M}{X} \frac{dX}{dM}\biggr|_{\Lambda={\rm const.}} \, .\label{eq:tildeetaX}
\end{equation}
In the fundamental theory, $\Lambda$ is constant and $\tilde \eta_X=\eta_X$, but in the effective theory
$\tilde \eta_X$ captures only the explicit mass dependence of the quantity in question and vanishes e.g. for purely gluonic
quantities.
It has been computed in~\cite{Cali:2019enm} for the masses\footnote{The quantities $\eta_{P,V}$ in~\cite{Cali:2019enm} for $\Nf=2$ correspond to
$\eta_{m_P,m_V}$ here, and for $\Nf=0$ correspond to $\tilde \eta_{m_P,m_V}$.} and found to deviate significantly from the
corresponding two-flavor values. Most of this deviation is explained by the neglected mass dependence associated with the scale. If the mass parameter
in the $\Nf=0$ theory is fixed by demanding that $m_P$ has the same value as in the $\Nf=2$ theory, the full mass dependence is
given by
\begin{equation}
        \eta_X^{(0)} = \eta^M + \frac{\tilde \eta_X^{(0)}}{\tilde \eta_{m_P}^{(0)}}\left(\eta_{m_P}^{(2)}-\eta^M \right) + O(M^{-2})\, .\label{eq:etaX0}
\end{equation}
In this equation, $\eta^M$ denotes the universal mass scaling function introduced in~\cite{Bruno:2014ufa}, which encodes the complete mass
dependence of purely gluonic quantities. Its definition is 
\begin{equation}
\eta^M = \frac{M}{P_{2,0}} \frac{\partial P_{2,0}}{\partial M}\bigg\vert_{\Lambda^{(2)}},
\end{equation}
where $P_{2,0}$ is defined in Eq.~\eqref{e:decoupling} and $M\equiv M^{(2)}$. A derivation of Eq.~(\ref{eq:etaX0}) is relegated to Appendix~\ref{sec:etaX}.

\section{Distance preconditioning for the Dirac operator}
\label{sec:distance_preconditioning}
The computation of $R_P$ and $R_{A_i}$ in Eqs.~\eqref{eq:rp} and
\eqref{eq:ra} requires the evaluation of the two-point correlation functions
$\bar f_{PP}(T-y_0,y_0)$ and $\bar f_{A_iA_i}(T-y_0,y_0)$.
These are constructed in the standard way in terms of quark propagators that
result from solving the Dirac equation $\sum_{x}D(y,x)\psi(x)=\eta(y)$, 
where $D(y,x)$ is the lattice Dirac operator, while $\psi(x)$ and $\eta(y)$
denote the solution vector and the source field (the latter being non-zero only on a
single time-slice $y_0$), respectively.
Whereas the increase of computational demand for the numerical inversion of
the Dirac operator to solve this linear system towards small quark masses
is a consequence of the spectral properties of $D(x,y)$, a difficulty of
entirely different nature arises in the region of heavy quark masses.

To understand the origin of this potentially serious problem, we recall that
the exponential decay with time of two-point functions at zero spatial
momentum is governed by the ground state energy in its spectral
decomposition, which equivalently reflects in the exponential time
dependence of the entering quark propagators proportional to
$\exp\{-m_q(x_0-y_0)\}$, where $m_q$ is the quark mass in the corresponding
Dirac operator.
However, the exponential decay of the quark propagator is particularly
pronounced in the case of heavy quarks, $m_q=m_h$, when the time distance
between source and sink grows. 
This entails that an accurate determination of heavy meson correlators over
the relevant time span is very difficult to achieve, because, depending on
the size of the heavy quark mass in lattice units, the relative precision of
the solution of the Dirac equation could actually start to deteriorate at
already moderately large time separations $x_0-y_0$. Moreover, this problem is amplified for correlation functions in the vector
channel, in which the ground state is even heavier.

In practice, for heavy quarks, the stopping criterion commonly imposed on
the relative \textit{global residuum}
\begin{equation}
	\frac{\bigl\lvert \sum\limits_{y}[D\psi(y)-\eta(y)]\bigr\rvert}{\bigl\lvert \sum\limits_{y} \eta(y)\bigr\rvert}
< r_{\rm gl}
\label{eq:global_residual}
\end{equation}
may therefore no longer appear as a reliable quantitative measure to
indicate the convergence of the iterative routine used to numerically solve
the Dirac equation to desired accuracy.
In fact, for large time separations $x_0-y_0$, contributions to the norm on
the l.h.s. suffer from a severe exponential suppression
$\propto\exp\{-m_h(x_0-y_0)\}$ and thus become negligible.
Even a ``brute force'' approach of reducing the global residual $r_{\rm gl}$
further to catch this may then not be feasible anymore for lattices with
very large time extents, regardless of the related increase in computational
cost.

To overcome this unwanted exponential suppression problem induced by the
heavy quark propagator, we employ the \textit{distance preconditioning (DP)}
technique for the Dirac operator, originally proposed in
Ref.~\cite{deDivitiis:2010ya}, in its implementation outlined
in~\cite{Collins:2017iud}. 
The basic idea is to rewrite the original linear system in such a way that
the solution of the new system at time-slices $x_0$ far away from the source
gets exponentially enhanced and thereby compensates the rapid decay of the
(heavy) quark propagator.
While in~\cite{deDivitiis:2010ya} the associated re-definition of the
propagator and source field was introduced on the level of the covariant
derivatives in the lattice Dirac operator, the DP implementation of
Ref.~\cite{Collins:2017iud} directly works with the matrix-vector equation
to be solved, since this has proven to be most straightforward for, e.g.,
the SAP-GCR solver routine~\cite{Luscher:2003qa} in use.
Schematically written, DP then amounts to replace
\begin{equation}
D\psi=\eta\quad\rightarrow\quad
D'\psi'=\eta',\quad D'=M D M^{-1},\quad \psi'=M\psi,\quad \eta'=M\eta,
\label{eq:DP}
\end{equation}
where in our setup the preconditioning matrix is defined as
\begin{equation}
M=M(x_0-y_0)=\mathrm{diag}\left(p_0,p_1,\ldots,p_{T/a}\right)
\quad\mathrm{with}\quad
p_i=\exp\left\{\alpha_0\big(x_0^{(i)}-y_0)\right\}\,.
\label{eq:DP_matrix}
\end{equation}
Here, $y_0$ and $x_0^{(i)}$ refer to the time-slices of the source and sink
insertion, and apart from this, $M$ is unity in spatial coordinates as well
as in spin and color spaces.
The solution of the original system is then simply obtained by
$\psi=M^{-1}\psi'$ from the solution of the preconditioned one.
In Eq.~\eqref{eq:DP_matrix}, $\alpha_0$ is a tunable parameter that
needs to be adjusted for each gauge field configuration ensemble in
dependence of the respective temporal lattice extent and heavy quark mass
in lattice units.

Rather than the global residual, Eq.~\eqref{eq:global_residual}, a more
suitable measure for the numerical quality of the solution over the whole
time range is now provided by inspecting the
\textit{local residuum},
\begin{equation}
r_{\rm loc}(x_0-y_0)=
\frac{\bigl\lvert\sum\limits_{\vec{y}}[D\psi(y)-\eta(y)]\bigr\rvert}
{\bigl\lvert\sum\limits_{\vec{y}}\psi(y)\bigr\rvert}\,,
\label{eq:local_residual}
\end{equation}
at the time separation of interest ($T-2y_0$ in our case).
Since $r_{\rm loc}$ is sensitive to the numerical accuracy of the solution on
each time-slice, imposing the stopping criterion of the routine involved
to solve the preconditioned system ensures to extract heavy meson
correlation functions to sufficient precision also for large time separations.
Note that the exponential factor $e^{\alpha_0(x_0-y_0)}$ in the construction
of the preconditioned linear system, Eq.~\eqref{eq:DP} above, effectively
turns the heavy quark propagator into a propagator corresponding to a
``ficticious'' light quark.
Therefore, the price to pay when adopting this method is an increase of the
number of solver iterations such that in practical applications
the growth in computational costs has to be reasonably balanced with
the gain in accuracy of the heavy meson correlator in question by a proper
tuning of $\alpha_0$.

\begin{figure}[h!]
\begin{minipage}{0.69\textwidth}
\includegraphics[width=0.99\textwidth]{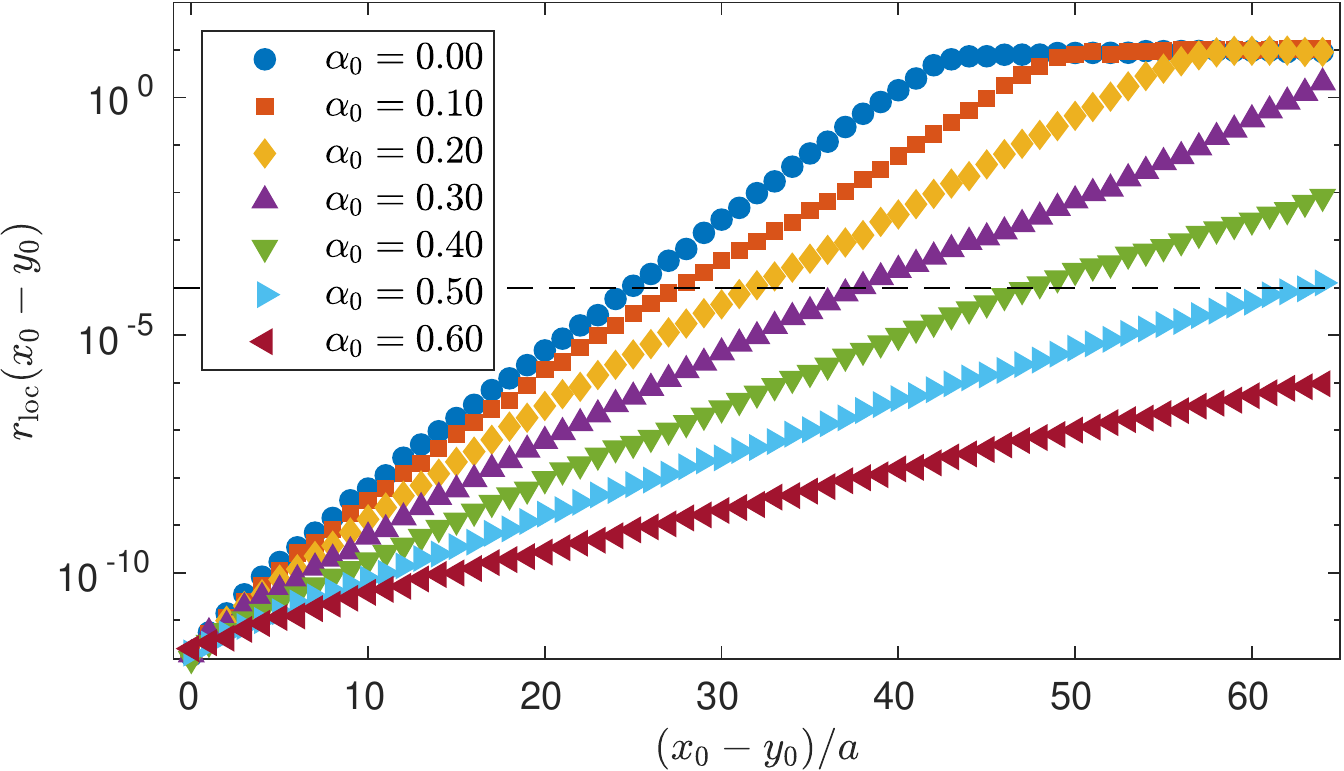}
\end{minipage}\hfill
\begin{minipage}{0.30\textwidth}
\begin{center}
\renewcommand{\arraystretch}{1.2}
\renewcommand{\tabcolsep}{6pt}
\begin{tabular}{c | c}
 $\alpha_0$ & Iterations\\
 \hline\hline
0.00 \hspace*{0.2cm} & 265\\
0.10 \hspace*{0.2cm} & 287\\
0.20 \hspace*{0.2cm} & 335\\
0.30 \hspace*{0.2cm} & 411\\
0.40 \hspace*{0.2cm} & 536\\
0.50 \hspace*{0.2cm} & 768\\
0.60 \hspace*{0.2cm} & 1296
\end{tabular}
\end{center}
\end{minipage}
\caption[Tuning of the parameter $\alpha_0$ of the distance preconditioning method for our coarsest lattice.]{Tuning of the parameter $\alpha_0$ of the distance preconditioning method for our coarsest lattice ($\Nf=2$, $\beta=5.3$, see Table~\ref{tab:ensembles})\
. We show the relative local residual of the solution $\psi$ as a function of the sink position $x_0$ with respect to the source set at $y_0/a=16$. On the right, the number of iterations for the solver to converge is reported for the values of $\alpha_0$ plotted in the Figure.}
\label{fig:tuning}
\end{figure}

As can be seen from monitoring $r_{\rm loc}$ for a representative ensemble and various choices of $\alpha_0$ in Figure~\ref{fig:tuning}, the solution becomes more and more accurate as $\alpha_0$ increases. However, in this example we observe that for the solution to have a local residual $< 10^{-4}$ for all time-slices (which is enough given the desired statistical accuracy of the meson correlator) the number of solver iterations increases of around a factor 3 compared to the case $\alpha_0=0$ (no distance preconditioning applied).
\noindent Such a small value of the residual is crucial to extract the meson correlator (and as a consequence the meson decay constants) reliably. Ideally one would choose $y_0 = a$, but to keep the computational effort as small as possible, we explored different source positions (with $y_0 > a$) to ensure the ground state dominance over a large range of time-slices and a reasonable number of iterations for the solver to converge.

Both $R_P$ and $R_{A_i}$ have been determined for all the ensembles under study and the parameters used in the solver setup are collected in Table~\ref{tab:parameters_dp}. 
\begin{table}[h!]
\centering
\begin{tabular}{c c c}
\toprule
Ensemble & $\alpha_0$ & $y_0/a$\\
\midrule
E        & 0.40 & 24   \\
N        & 0.10 & 30    \\
P        & 0.05 & 20    \\
S        & 0.01 & 48    \\
W        & 0.02 & 32    \\
\midrule
qN       & 0.10 & 30  \\
qP       & 0.05 & 20  \\
qW       & 0.02 & 32  \\
qX       & 0.00 & 32  \\
\bottomrule
\end{tabular}
\caption{The table lists the values of the parameter $\alpha_0$ for the distance preconditioning of the Dirac operator and the time-slice of the source position used to determine $R_P$ and $R_{A_i}$. Distance preconditioning is not used for the ensemble qX.}\label{tab:parameters_dp}
\end{table}

\section{Results}\label{sec:results}

\subsection{Decay constants of charmonium}

Combining the measurements of the ratios $R_P$ and $R_{A_i}$, defined in Eqs.~\eqref{eq:rp} and~\eqref{eq:ra}, with the charmonium masses that we previously obtained in Ref.~\cite{Cali:2019enm}, through Eqs.~\eqref{eq:fp_formula} and~\eqref{eq:fv_formula}, we determine the dimensionless quantities $\sqrt{t_0}\fP$ and $\sqrt{t_0}\fV$. 
They are listed in Appendix~\ref{s:tabledecay}.

\begin{figure}[t]
\begin{center}
\includegraphics[width=0.7\textwidth]{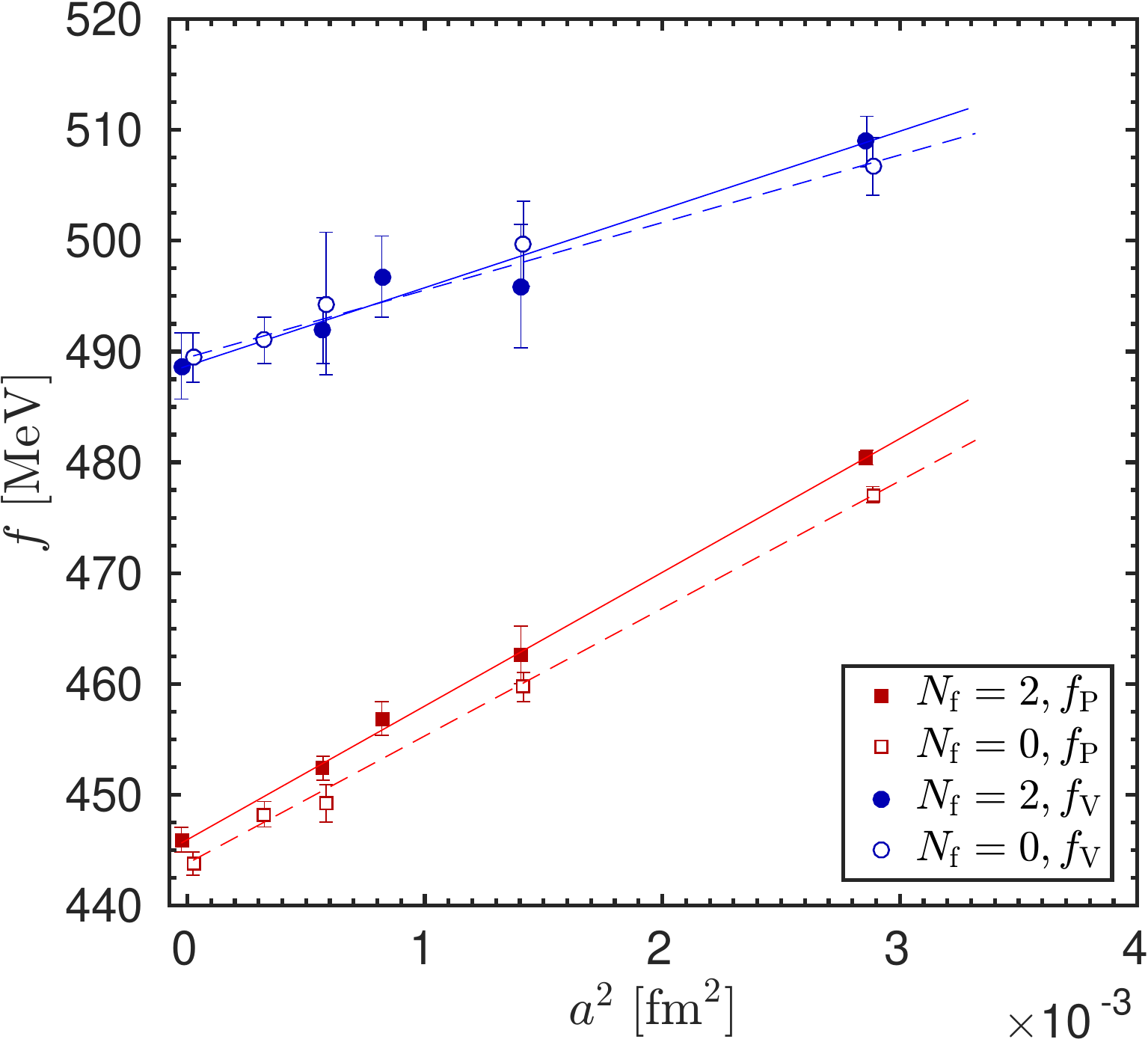}
\caption{Continuum extrapolations of the meson decay constants $\fP$ and $\fV$ on our $\Nf=2$ (full markers) and $\Nf=0$ (empty markers) ensembles. The continuum limit values are slightly displaced horizontally. To set the scale in physical units, we use $\sqrt{t_0}=0.1131(38)$~fm~\cite{Athenodorou:2018wpk}.}\label{fig:decay_constants}
\end{center}
\end{figure}

Similarly to what we observed for the meson masses in Ref.~\cite{Cali:2019enm}, we find that linear fits in $a^2$ do not work well in the range of lattice spacings $0.02\mbox{ fm}\lesssim a \lesssim 0.07\mbox{ fm}$, if one aims at results with sub-percent precision.
Therefore, we compare the continuum limits of pseudoscalar and vector decay constants in $\Nf = 0$ and $\Nf = 2$ QCD, performing extrapolations to zero lattice spacing only in the range $0.02\mbox{ fm}\lesssim a \lesssim 0.05\mbox{ fm}$. The results for the $\Nf=0$ (empty markers) and $\Nf=2$ (full markers) ensembles are shown in Figure~\ref{fig:decay_constants}. The continuum limit values are slightly displaced for clarity.

Comparing the values obtained in the continuum of the two theories, one can observe that dynamical charm effects on the decay constants are barely resolvable, despite the great accuracy of our continuum extrapolations. In Table~\ref{tab:contextrap} we summarize our findings.
The relative difference of $\sqrt{t_0}\fP$ is $([\sqrt{t_0}\fP]^{\Nf=2}-[\sqrt{t_0}\fP]^{\Nf=0})/[\sqrt{t_0}\fP]^{\Nf=2}=0.48(34)\%$, which corresponds to an effect of around $1.4\sigma$.
\begin{table}[h!]
\centering
\begin{tabular}{l l l l}
\toprule
Quantity        & $\Nf=2$     & $\Nf=0$     & sea effects [\%] \\
\midrule
$\sqrt{t_0}\fP$ & 0.25604(65) & 0.25481(59) & 0.48(34)\\
$\sqrt{t_0}\fV$ & 0.2806(17)  & 0.2810(13)  & 0.12(77)\\
\bottomrule
\end{tabular}
\caption{Results for $\sqrt{t_0}\fP$ and $\sqrt{t_0}\fV$ in the continuum limit for both the $\Nf=2$ and the $\Nf=0$ theory.}\label{tab:contextrap}
\end{table}
Moreover, employing $\sqrt{t_0}=0.1131(38)$~fm~\cite{Athenodorou:2018wpk} for our model of QCD, we obtain\footnote{The $\Nf=2$ result~\eqref{eq:f_nf2_phys_units} is obtained in QCD with two dynamical charm quarks but no light quarks. The $\Nf=0$ result~\eqref{eq:f_nf0_phys_units} relies on decoupling to set the scale through the value of $t_0$ determined in the $\Nf=2$ theory.}
\begin{eqnarray}
&\fP&= 445.9(1.1)\text{ MeV},\quad \fV= 488.7(3.0)\text{ MeV}\qquad (\Nf=2),\label{eq:f_nf2_phys_units}\\
&\fP&= 443.8(1.0)\text{ MeV},\quad \fV= 489.4(2.2)\text{ MeV}\qquad (\Nf=0)\label{eq:f_nf0_phys_units}.
\end{eqnarray}
Note that the decay constant of the meson $\eta_c$ has not been determined experimentally, whilst for the vector meson $J/\psi$ the value $f_{J/\psi}=407(4)$ MeV is obtained from the partial decay width of $J/\psi$ into an electron-positron pair, see \cite{Hatton:2020qhk}. $f_{J/\psi}$ can be compared to our lattice results for $\fV$ in Eqs.~\eqref{eq:f_nf2_phys_units} and~\eqref{eq:f_nf0_phys_units}. The discrepancy is probably due to effects of light sea quarks, disconnected contributions and electromagnetism, which are neglected in this work, and also to the unphysical values of the meson masses.
\begin{figure}[h!]
\centering
\includegraphics[width=0.7\linewidth]{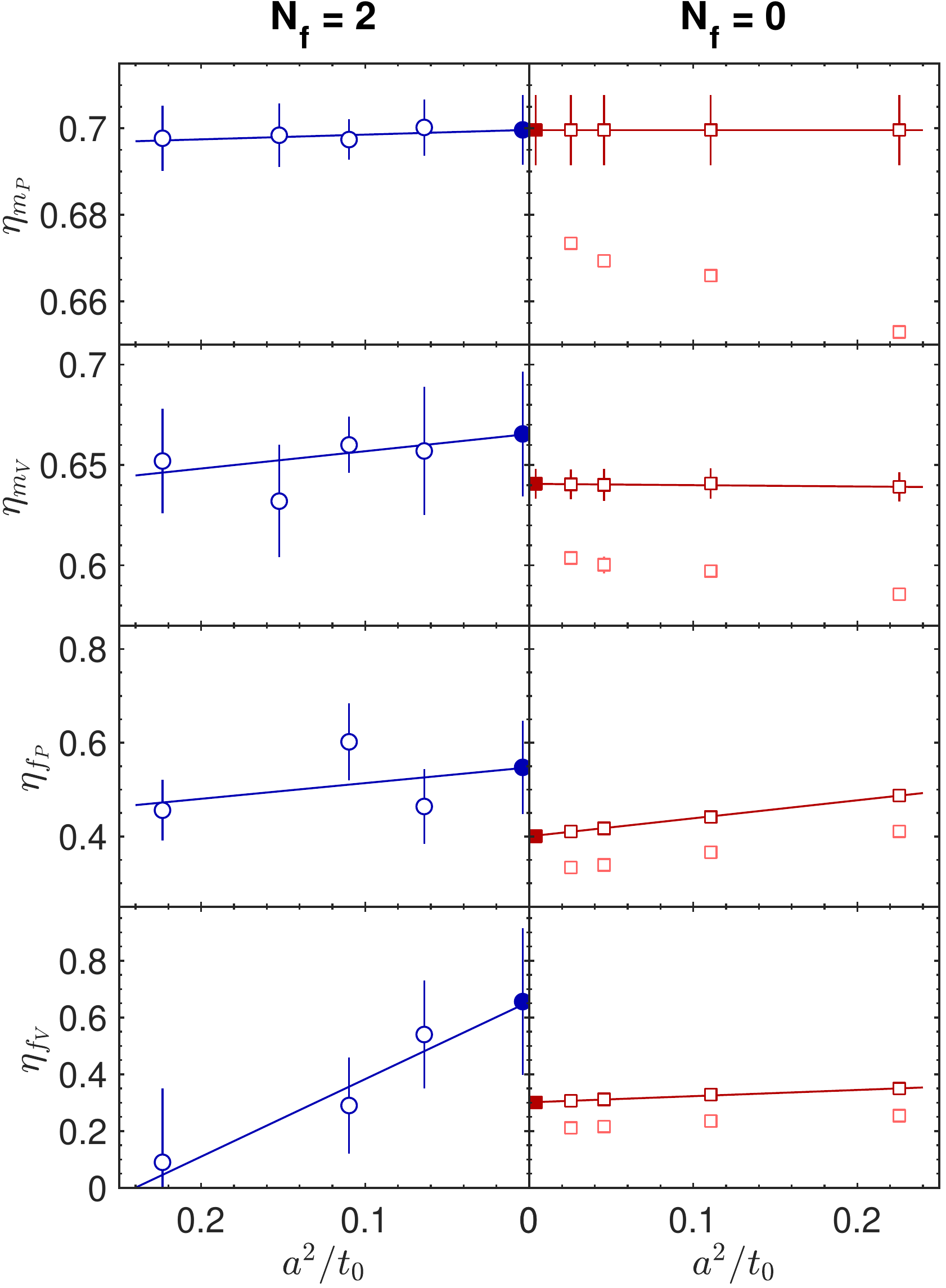}\\
	\caption{Logarithmic derivative of the pseudoscalar and vector masses and decay constants with respect to the quark mass. Comparison between $\Nf=2$ and $\Nf=0$ theories. For the $\Nf=0$ case, the values of $\tilde \eta_X$, where
	the mass dependence of the scale is neglected, are shown in pink. When the scale dependence is taken into account, the errors 
	increase due to a dependence on $\eta_{m_P}^{(2)}$.}\label{fig:deriv_F}
\end{figure}

The mass dependence functions for meson masses and decay constants have been computed as well.
Figure~\ref{fig:deriv_F} shows how $\eta_X^{(2)}$ compares to both $\eta_X^{(0)}$ and $\tilde \eta_X^{(0)}$ for various quantities. 
In $\eta_X^{(2)}$ and $\tilde \eta_X^{(0)}$, the derivatives are computed using Eqs.~\eqref{eq:dfdmu} and~\eqref{eq:dAdmu}. The missing pieces to evaluate Eq.~(\ref{eq:etaX0}) are the continuum value $\eta_{m_P}^{(2)}=0.6996(81)$ from~\cite{Cali:2019enm}, and the value
$\eta^M(M_c)=0.1276(2)$ determined as in~\cite{Athenodorou:2018wpk} for the $\Nf=2 \to \Nf=0$ case, 
using four- and five-loop perturbation theory results for the
decoupling relations~\cite{Chetyrkin:2005ia,Schroder:2005hy}, the QCD $\beta$-function and the QCD anomalous dimension~\cite{vanRitbergen:1997va,Czakon:2004bu,Baikov:2016tgj,Luthe:2016ima,Herzog:2017ohr}. Table~\ref{tab:contextrap_der}
summarizes the findings on the mass dependence. Note that agreement is only observed when the mass dependence is consistently taken into account.

\begin{table}[t]
\centering
\begin{tabular}{l l l l}
\toprule
$X$       &  $\eta^{(2)}_X$ & $\eta^{(0)}_X$ & $\tilde \eta^{(0)}_X$ \\
\midrule
$m_{P}$   & 0.6996(81)      & 0.6996(81)     & 0.67553(42)           \\
$m_{V}$   & 0.665(31)       & 0.6406(73)     & 0.6060(13)            \\
$f_{P}$   & 0.547(99)       & 0.4003(43)     & 0.3233(19)            \\
$f_{V}$   & 0.66(26)        & 0.3010(74)     & 0.2058(81)            \\
\bottomrule
\end{tabular}
\caption{The mass dependence functions for various quantities in the continuum limit, at the charm quark mass. The results for the meson masses rely on data published in~\cite{Cali:2019enm}. Note that in the first row $\eta_{m_P}^{(2)}=\eta_{m_P}^{(0)}$ because of the matching condition~\eqref{e:matching}.\label{tab:contextrap_der}}
\end{table}



The charm sea quarks effects in the mass derivatives of an observable
  originate from the covariance of the
  mass derivative of the effective action with the observables, see Eq.~\eqref{eq:dAdmu_gauge}.
  This covariance is absent in the $\Nf=0$ theory and explains the much larger
  statistical errors of the $\Nf=2$ theory.
  The derivatives computed on each ensemble are used to shift the decay constants to the tuning point, see Eq.~\eqref{eq:phistar}. Still, they are physical quantities on their own.
  
\subsection{Heavy meson masses}\label{subsec:hc_mesons}
In this section we present our results for heavy mesons made up of a heavy quark $h$ ($M_h=2M_c$) and a charm quark $c$, as described in Section~\ref{sec:obs_heavy_mesons}. To guarantee that the cutoff effects in the continuum extrapolations are under control, we only consider the ensembles of Table~\ref{tab:ensembles} for which the meson masses satisfy the relation $am<1$.
\begin{figure}[h!]
\centering
\includegraphics[width=0.8\linewidth]{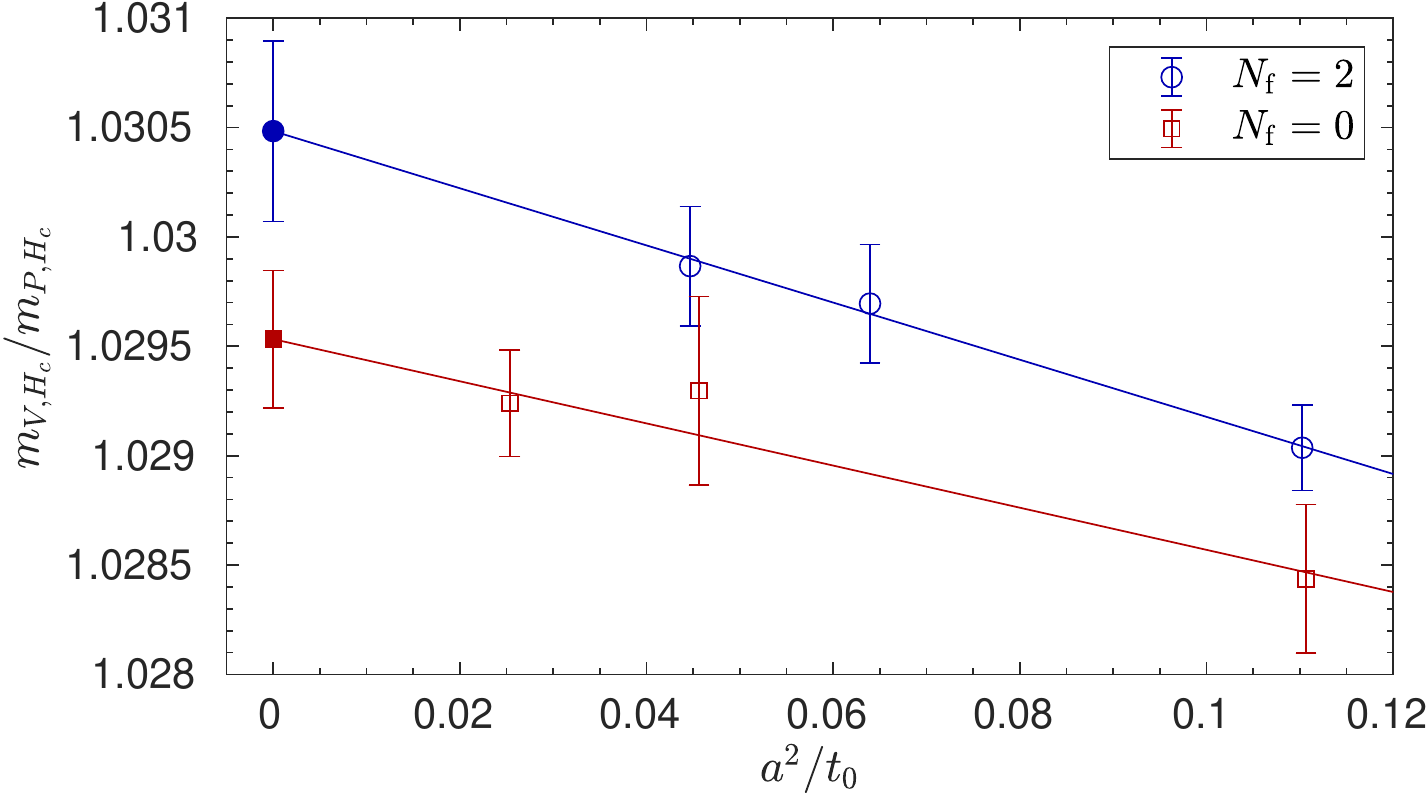}
 \caption{Continuum limits of the ratio $m_{V,H_c}/m_{P,H_c}$ for $\Nf = 2$ (blue circles) and $\Nf=0$ (red squares) QCD. The mass of the heavy quark is $M_h=2M_c$.}\label{fig:ratio}
\end{figure}

In Figure~\ref{fig:ratio} we compare the continuum limits of the mass ratio $m_{V,H_c}/m_{P,H_c}$ in $\Nf = 0$ and $\Nf = 2$ QCD. The numerical results of the continuum extrapolations are summarized in the first row of Table~\ref{tab:contextrap1}.
\begin{table}[h!]
\centering
\begin{tabular}{l l l l}
\toprule
Quantity        & $\Nf=2$     & $\Nf=0$     & sea effects [\%] \\
\midrule
$m_{V,H_c}/m_{P,H_c}$, $M_h=2M_c$ & 1.03048(41) & 1.02953(31) & 0.092(50)\\
$m_{V,H_c}/m_{P,H_c}$, $M_h=M_c$  & 1.05405(60) & 1.05274(46) & 0.124(71)\\
\bottomrule
\end{tabular}
\caption{Results for $m_{V,H_c}/m_{P,H_c}$ in the continuum limit for both the 
$\Nf=0$ and the $\Nf=2$ theory. The values for the $M_h=M_c$ case are taken from Ref.~\cite{Cali:2019enm}.}\label{tab:contextrap1}
\end{table}
As it can be seen, the effect of dynamical charm quarks on the ratio $m_{V,H_c}/m_{P,H_c}$ is found to be $0.092(50)\%$ (around $1.8\sigma$), which is of similar size to the one obtained in our previous work \cite{Cali:2019enm} for $M_h=M_c$. 

\begin{figure}[h!]
\centering
\includegraphics[width=0.8\linewidth]{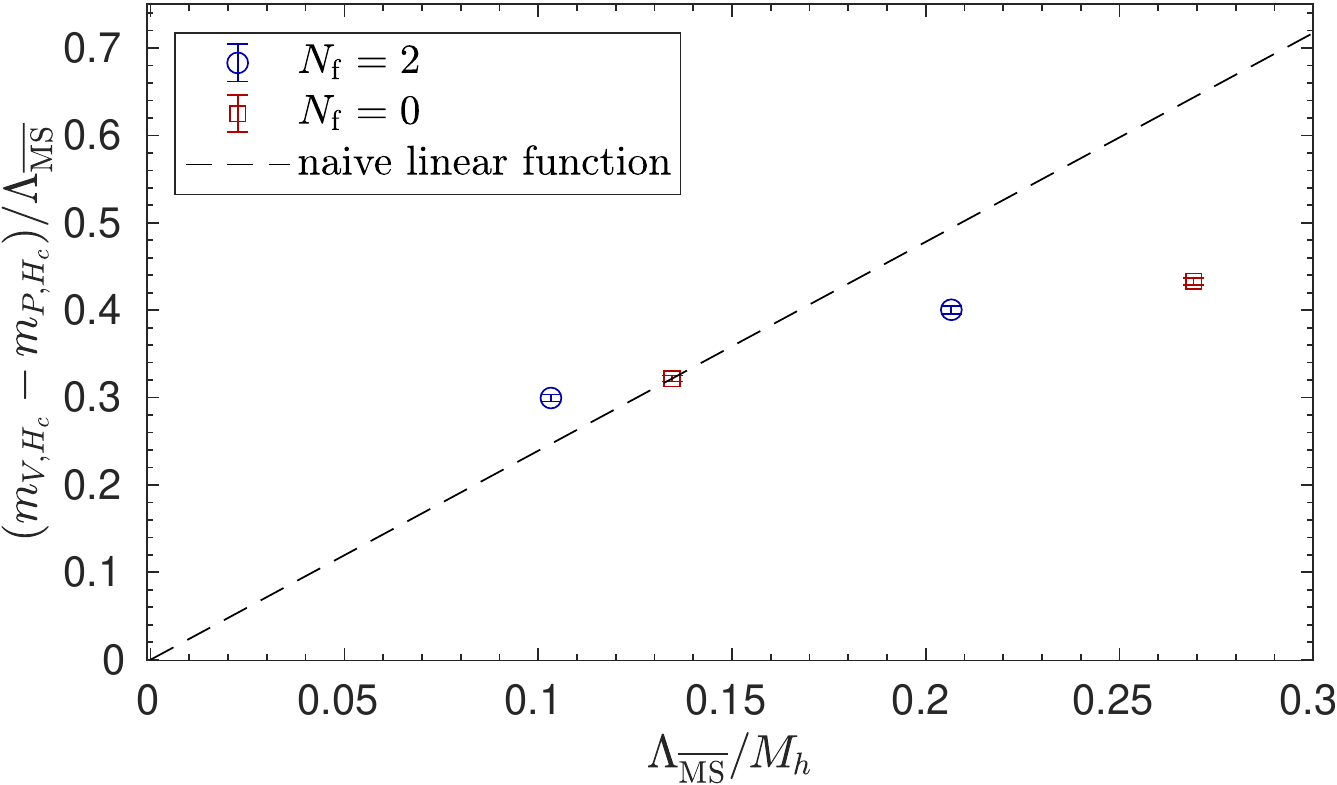}
 \caption{The hyperfine splitting of heavy-charm mesons as a function of the inverse RGI heavy quark mass for $\Nf = 2$ (blue circles) and $\Nf=0$ (red squares) QCD. The dashed line is a linear function which passes through the $\Nf=0$ points at infinite quark mass and $M_{h}=2M_c$, and it has been extended to smaller quark masses.}\label{fig:staticlimit}
\end{figure}

In Fig.~\ref{fig:staticlimit} we show the hyperfine splitting $(m_{V,H_c}-m_{P,H_c})/\Lambda_{\msbar}$ as a function of $\Lambda_{\msbar}/M_{h}$, where $M_{h}$ is the RGI mass of the heavy quark $h$. The blue circles correspond to the $\Nf=2$ theory and the red squares to the pure gauge theory, $\Nf=0$. We take the $\Lambda$ parameters in units of the scale $\sqrt{t_0}$ \cite{Luscher:2010iy} from \cite{Fritzsch:2012wq,Athenodorou:2018wpk} for $\Nf=2$ QCD and from \cite{DallaBrida:2019wur} for $\Nf=0$ QCD. The RGI quark mass values $M_h=M_c$ at our tuning point Eq.~\eqref{eq:tuning_point} have been determined in \cite{Cali:2019enm} for both theories. The data of Figure~\ref{fig:staticlimit} are listed in Table~\ref{tab:hyperfine}.

  The limit $M_{h}\to\infty$ is expected to be described by Heavy Quark Effective Theory (HQET). For $\Nf=0$ QCD we show in Figure~\ref{fig:staticlimit} a linear function (dashed line) going through the points at $M_{h}=\infty$, where the vector and the pseudoscalar are degenerate by virtue of the heavy quarks spin symmetry \cite{Isgur:1989vq,Isgur:1989ed,Sommer:2010ic}, and $M_{h}=2M_c$. Contact with HQET in the spirit of~\cite{Heitger:2004gb}, including the HQET-QCD matching factor $C_\mathrm{spin}$ that accounts for the correct mass dependence of the $M_h\rightarrow\infty$ asymptotics, would require larger heavy quark masses than we have simulated in this work.

\begin{table}[h!]
\centering
\begin{tabular}{l| l l| l l}
\toprule
$M_{h}$ & \multicolumn{2}{c|}{$\Nf=2$}     & \multicolumn{2}{c}{$\Nf=0$}     \\
& $\Lambda_{\msbar}/M_{h}$ & $(m_{V,H_c}-m_{P,H_c})/\Lambda_{\msbar}$ &
$\Lambda_{\msbar}/M_{h}$ & $(m_{V,H_c}-m_{P,H_c})/\Lambda_{\msbar}$ \\
\midrule
$2M_c$ & 0.1033 & 0.2995(40) & 0.1346 & 0.3217(34) \\
$M_c$  & 0.2066 & 0.4005(45) & 0.2691 & 0.4330(38) \\
\bottomrule
\end{tabular}
\caption{Hyperfine splitting of heavy-charm mesons
    $(m_{V,H_c}-m_{P,H_c})/\Lambda_{\msbar}$ as a function of $\Lambda/M_{h}$,
    where $M_{h}$ is the RGI mass of the heavy quark $h$.}
\label{tab:hyperfine}
\end{table}

\section{Conclusions}

In this work we present a study of the effects of sea charm quarks in the charmonium decay constants. It is a follow-up of Ref.~\cite{Cali:2019enm} where we did a similar study for the charmonium spectrum and the RGI charm quark mass. We compute the charm-quark sea effects through a comparison of the decay constants determined in a model of QCD with only $\Nf=2$ charm quarks and no light quarks with those computed in the Yang--Mills theory ($\Nf=0$ QCD). The charm quark mass is fixed in both theories by the tuning condition Eq.~\eqref{eq:tuning_point}, which approximately corresponds to a physical charm quark.

  We find that the effects of two charm sea quarks are below 1\% for the decay constant of the pseudoscalar ($\eta_c$) and 
of the vector ($J/\psi$) ground state particles. We also extracted the derivatives of the decay constants and meson masses with respect to the charm quark mass, 
because they are required to shift our simulation results to the tuning point Eq.~\eqref{eq:tuning_point}. 
The quantities $\eta_X = \frac{M}{X}\frac{d X}{dM}$ ($X=m_P,m_V,f_P,f_V$) are renormalized and easily computable in the twisted mass formulation at maximal twist, because the twisted mass parameter $\mu$ is multiplicatively renormalizable. In the continuum limit we do not see
significant effects of the dynamical charm quark on these quantities.

  A computation of the charm sea effects for ``$B_c$'' mesons made of a charm quark and an antiquark of twice larger mass reveals that they are only about $1${\textperthousand} for the ratio of the vector to the pseudoscalar mass.

  Our results are relevant for the scale setting \cite{Bruno:2016plf} of the $\Nf=2+1$ simulations by the CLS consortium \cite{Bruno:2014jqa,Bali:2016umi}. There, a combination of the pseudoscalar decay constants of the pion and kaon is used. They are lower in energy than the decay constants of charmonium. Hence our results indicate that charm sea effects are below the precision of the scale determination in \cite{Bruno:2016plf}. Finally, the analysis of the ``$B_c$'' mesons also demonstrates that charm sea effects are very small for splittings of heavy mesons. For instance, the $\Upsilon-\Upsilon'$ splitting has been used in \cite{Davies:2003ik} for scale setting.

\section*{Acknowledgments}
We thank Rainer Sommer for his useful remarks on the mass dependence functions. The authors gratefully acknowledge the Gauss Centre for Supercomputing e.V. (\url{www.gauss-centre.eu}) for funding this project by providing computing time on the GCS Supercomputer SuperMUC-NG at Leibniz Supercomputing Centre (\url{www.lrz.de}). S.C. acknowledges support from the European Union's Horizon 2020 research and innovation programme under the Marie Sk\l{}odowska-Curie grant agreement No. 642069, the polish NCN grant No. 2016/21/B/ST2/01492 and the Carl G and Shirley Sontheimer Research Fund at MIT.
This work is also supported by the Deutsche Forschungsgemeinschaft (DFG) through the Research Training Group ``GRK 2149: Strong and Weak Interactions -- from Hadrons to Dark Matter'' (K.E. and J.H.). 

\appendix
\section{Table of decay constants}
\label{s:tabledecay}

All the results obtained on our ensembles are listed in Table~\ref{tab:decay_consts}.
\begin{table}[h!]
\centering
\begin{tabular}{c l l l l l l l l}
\toprule
Ensemble & $a\mu$ & $\sqrt{t_0}\fP$ & $\sqrt{t_0}\fV$ &
$\frac{\mu}{f_P}\frac{d f_P}{d\mu}$ & $\frac{\mu}{f_V}\frac{d f_V}{d\mu}$ \\
\midrule
N        & 0.16647(28) & 0.27582(38) & 0.2922(13) &&\\
         & 0.166 & 0.27553(41) &  0.2923(13) & 0.456(65) & 0.09(26) \\
P        & 0.11482(32) & 0.2656(15) & 0.2847(32) &&\\
         & 0.1132 & 0.2637(16) &  0.2840(36) & 0.602(82) & 0.29(17) \\
S        & 0.08717(25) & 0.26229(87) & 0.2852(21) &&\\
         & 0.087626 & 0.26291(84) & 0.2860(18) & 0.464(80) & 0.54(19)\\
W        & 0.072557 & 0.25972(62) & 0.2824(17) & -- & -- \\
\midrule
qN        & 0.17632(11) & 0.27390(42) &  0.2909(15) & 0.4109(18) & 0.254(15)\\
          & 0.16 & 0.26314(47) & 0.2838(17) &&\\
          & 0.17 & 0.26967(47) & 0.2881(16) && \\
          & 0.18 & 0.27618(47) & 0.2924(14) && \\
qP        & 0.12235(26) & 0.26392(75) & 0.2869(22) & 0.3663(28) & 0.235(22)\\
          & 0.11 & 0.25394(78) & 0.2799(27) && \\
          & 0.12 & 0.26208(77) & 0.2857(23) && \\
          & 0.13 & 0.26994(76) & 0.2911(20) && \\
qW        & 0.07798(19) & 0.25790(97) & 0.2838(37) & 0.3393(34) &  0.216(24)\\
          & 0.07 & 0.2494(11) & 0.2779(40) &&\\
          & 0.08 & 0.2607(11) & 0.2857(35) &&\\
          & 0.09 & 0.2715(11) & 0.2933(30) &&\\
qX        & 0.05771(13) & 0.25735(66) & 0.2819(12) & 0.3336(20) & 0.2110(73)\\
          & 0.056 & 0.25480(69) & 0.2801(13) &&\\
          & 0.058 & 0.25780(68) & 0.2822(12) &&\\
          & 0.06  & 0.26073(68) & 0.2842(12) &&\\
\bottomrule
\end{tabular}
\caption{Meson decay constants in $t_0$ units. For $\Nf=2$ simulations,  the first line contains the values extrapolated to the tuning point $\mu^\star$ and the second line
  the values at the simulated parameters. For $\Nf=0$ ensembles, the first line gives the values interpolated to $\mu^\star$, and the following three lines contain the values measured at different valence quark masses. The last two columns are the values of the logarithmic derivatives of the decay constants with respect to the logarithm of the quark mass. For the ensemble W they have not been computed.}\label{tab:decay_consts}
\end{table}

\section{Mass dependence of fermionic observables}\label{sec:etaX}
Eq.~(\ref{eq:etaX0}) describes how quantities with an explicit valence quark mass dependence depend on the quark mass, if decoupling is 
in place. The parameters of the leading order effective theory, here $\Nf=0$ QCD, are fixed by demanding non-perturbatively
\begin{eqnarray}
   \sqrt{t_0^{(2)}}          &=& \sqrt{t_0^{(0)}}\, , \\
   \sqrt{t_0^{(2)}}m_P^{(2)} &=& \sqrt{t_0^{(0)}} m_P^{(0)}\, .
   \label{e:matching}
\end{eqnarray}
The first condition fixes the scale
\begin{equation}
  \Lambda^{(0)} = P_{2,0}(M^{(2)}/\Lambda^{(2)}) \ \Lambda^{(2)}\, , \label{e:decoupling}
\end{equation}
where
$P_{2,0}$ is a matching function that can also be computed perturbatively, and which is unique up to $O(M^{-2})$
power corrections~\cite{Bruno:2014ufa}. The second condition fixes the valence quark mass $M^{(0)}$ in the quenched theory.
With just two dimensionful parameters in the game, $M$ and $\Lambda$, every quantity of mass dimension 
one can be factorized into
\begin{equation}
   X^{(\Nf)}(M^{(\Nf)},\Lambda^{(\Nf)})  = f_X^{(\Nf)}(M^{(\Nf)}/\Lambda^{(\Nf)}) \ \Lambda^{(\Nf)}\, ,
\end{equation}
with some dimensionless function $f_X$. In the fundamental $\Nf=2$ theory, $\Lambda$ is constant, in the effective theory it depends
on $M^{(2)}\equiv M$ through the decoupling relation \eq{e:decoupling}. Moreover, $f_X$ is constant in the effective theory for all quantities that do not depend
on the valence quark mass $M^{(0)}$, e.g. all gluonic observables. Using the pseudoscalar meson mass to fix the $\Nf=0$ valence quark mass means that
it is given by
\begin{equation}
   M^{(0)} = {f_{m_P}^{(0)}}^{-1}\left(\frac{f_{m_P}^{(2)}(M/\Lambda^{(2)})}{P_{2,0}(M/\Lambda^{(2)})} \right) P_{2,0}(M/\Lambda^{(2)})\ \Lambda^{(2)}\, .\label{eq:M0}
\end{equation}
With this we can now evaluate
\begin{eqnarray}
        \eta_X^{(0)} &=& \frac{M}{X^{(0)}}\ \frac{d X^{(0)}}{d M} \\
		     &=& \frac{M}{X^{(0)}} \frac{d f_X^{(0)}(M^{(0)}/\Lambda^{(0)})\Lambda^{(0)}}{dM}\, .
\end{eqnarray}
Both $\Lambda^{(0)}$ and $M^{(0)}$ depend on $M$, as outlined above. Inserting this dependence into the equation and carrying out the derivatives using the
usual rules for derivatives of inverse functions, one finds
\begin{equation}
        \eta_X^{(0)} = \eta^M + \frac{\tilde \eta_X^{(0)}}{\tilde \eta_{m_P}^{(0)}}\left(\eta_{m_P}^{(2)}-\eta^M \right) + O(M^{-2})\, ,
\end{equation}
where $\tilde \eta^{(0)}_X = \frac{M^{(0)}}{X}\frac{dX}{dM^{(0)}}\bigr|_{\Lambda={\rm const}}$. The latter captures the valence quark mass dependence,
but ignores the mass dependence of the scale in the effective theory.

\vskip 0.3cm

\noindent
\clearpage
\bibliographystyle{UTjhep} 
\bibliography{biblio}

\end{document}